\documentclass[sigconf]{acmart}

\AtBeginDocument{%
  }

\setcopyright{rightsretained}
\copyrightyear{2024}
\acmYear{2024}

\acmConference[MuC'24]{Mensch und Computer 2024 – Workshopband, Gesellschaft für Informatik e.V.}{01.-04. September 2024}{Karlsruhe, Germany}
\acmDOI{10.18420/muc2024-mci-demo-303}
\acmPrice{}
\acmISBN{}

\usepackage[nolist,nohyperlinks, printonlyused, withpage]{acronym}
\usepackage{csquotes}
\usepackage{siunitx} 
\usepackage{tabularx}
\usepackage[inline]{enumitem} 
\usepackage{subfig}
\usepackage{amsmath}

\begin{acronym}
\acro{ARCS}{attention, relevance, confidence, satisfaction}
\acro{CEGE}{Core Elements of game experience}
\acro{DDA}{Dynamic Difficulty Adjustment}
\acro{FUGA}{Fun of gaming}
\acro{FPS}{First-person shooter}
\acro{GUR}{Games User Research}
\acro{HCI}{Human-Computer Interaction}
\acro{KIT}{Karlsruhe Institute of Technology}
\acro{MMO}{Massively Multiplayer Online}
\acro{PC}{Personal Computer}
\acro{PE}{Player Engagement}
\acro{PX}{Player Experience}
\acro{PXI}{Player Experience Inventory}
\acro{RPG}{Role-playing Game}
\acro{SDT}{Self-Determination Theory}
\acro{TAM}{Technology Acceptance Model}
\acro{UE}{User Engagement}
\acro{UX}{User Experience}
\acro{VR}{Virtual Reality}
\end{acronym}

\makeatletter
\gdef\@copyrightpermission{
	\begin{minipage}{0.2\columnwidth}
		\href{https://creativecommons.org/licenses/by/4.0/}{\includegraphics[width=0.90\textwidth]{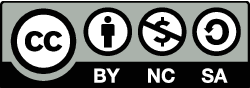}}
	\end{minipage}\hfill
	\begin{minipage}{0.8\columnwidth}
		\href{https://creativecommons.org/licenses/by/4.0/}{This work is licensed under a Creative Commons Attribution International 4.0 License.}
	\end{minipage}
	\vspace{5pt}
}
\makeatother

\begin{document}

\title[VR Simulation of Access Barriers]{Leveraging Virtual Reality Simulation to Engage Non-Disabled People in Reflection on Access Barriers for Disabled People}

\author{Timo Brogle}
\authornote{Authors contributed equally to this research.}
\orcid{0009-0004-2709-1704}
\email{timo.brogle@t-online.de}
\affiliation{%
  \institution{HCI and Accessibility, Karlsruhe Institute of Technology (KIT)}
  \city{Karlsruhe}  
  \country{Germany}
}

\author{Andrej Vladimirovic Ermoshkin}
\authornotemark[1]
\orcid{0009-0000-4764-9219}
\email{andrej.ermoshkin2004@gmail.com}
\affiliation{%
  \institution{HCI and Accessibility, Karlsruhe Institute of Technology (KIT)}
  \city{Karlsruhe}  
  \country{Germany}
}

\author{Konstantin Vakhutinskiy}
\authornotemark[1]
\orcid{0009-0000-9583-0792}
\email{konstvakh1@gmail.com}
\affiliation{%
  \institution{HCI and Accessibility, Karlsruhe Institute of Technology (KIT)}
  \city{Karlsruhe}  
  \country{Germany}
}

\author{Sven Priewe}
\authornotemark[1]
\orcid{0009-0000-2864-0358}
\email{sven.priewe@gmail.com}
\affiliation{%
  \institution{HCI and Accessibility, Karlsruhe Institute of Technology (KIT)}
  \city{Karlsruhe}  
  \country{Germany}
}

\author{Claas Wittig}
\authornotemark[1]
\orcid{0009-0001-7137-2333}
\email{wittigclaas@gmail.com}
\affiliation{%
  \institution{HCI and Accessibility, Karlsruhe Institute of Technology (KIT)}
  \city{Karlsruhe}  
  \country{Germany}
}

\author{Anna-Lena Meiners}
\orcid{0000-0002-9803-1555}
\email{anna-lena.meiners@kit.edu}
\affiliation{%
  \institution{HCI and Accessibility, Karlsruhe Institute of Technology (KIT)}
  \city{Karlsruhe}  
  \country{Germany}
}

\author{Kathrin Gerling}
\orcid{0000-0002-8449-6124}
\email{kathrin.gerling@kit.edu}
\affiliation{%
  \institution{HCI and Accessibility, Karlsruhe Institute of Technology (KIT)}
  \city{Karlsruhe}  
  \country{Germany}
}

\author{Dmitry Alexandrovsky}
\orcid{0000-0001-9551-719X}
\email{dmitry.alexandrovsky@kit.edu}
\affiliation{%
  \institution{HCI and Accessibility, Karlsruhe Institute of Technology (KIT)}
  \city{Karlsruhe}  
  \country{Germany}
}

\renewcommand{\shortauthors}{Brogle et al.}

\begin{abstract}
Disabled people experience many barriers in daily life, but non-disabled people rarely pause to reflect and engage in joint action to advocate for access. 
In this demo, we explore the potential of Virtual Reality (VR) to sensitize non-disabled people to barriers in the built environment. We contribute a VR simulation of a major traffic hub in Karlsruhe, Germany, and we employ visual embellishments and animations to showcase barriers and potential removal strategies. 
Through our work, we seek to engage users in conversation on what kind of environment is accessible to whom, and what equitable participation in society requires. Additionally, we aim to expand the understanding of how VR technology can promote reflection through interactive exploration.
\end{abstract}



\begin{teaserfigure}
\centering
\subfloat[\Description{Screenshot of the VR simulation. In view is the sidewalk with a tactile guiding strip in the center, framed by grass on the sides. In the distance, there is a road crossing the sidewalk with cars driving along, and there is other infrastructure, e.g., a church, and buildings related to the tram and subway stations.}]{\includegraphics[width=0.3\textwidth]{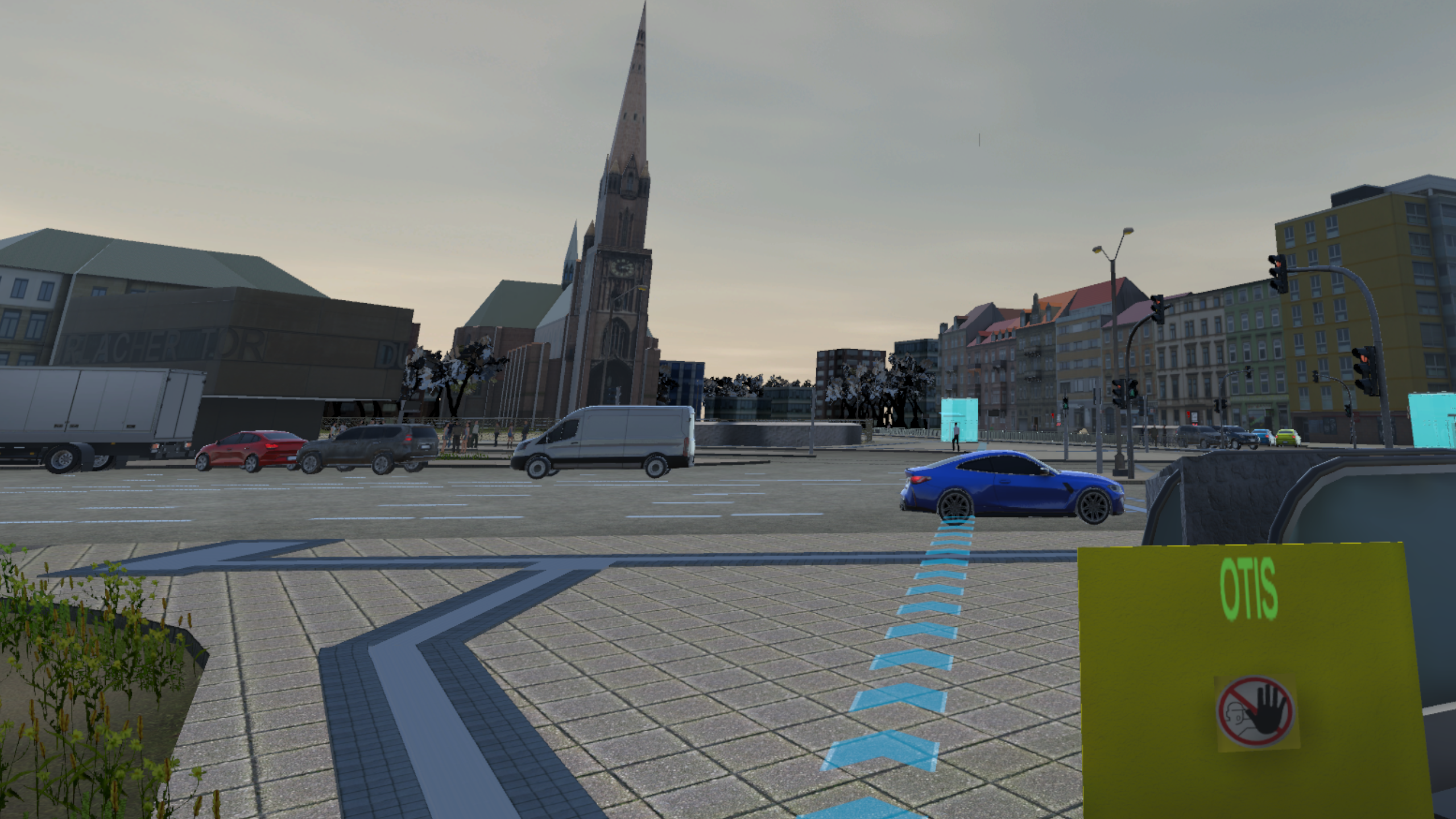}}
\qquad
\subfloat[\Description{Screenshot of the VR simulation. In view is the tram station. In front is a tram arriving. On the sidewalk to the left and right, groups of pedestrians are waiting at the station.}]{\includegraphics[width=0.3\textwidth]{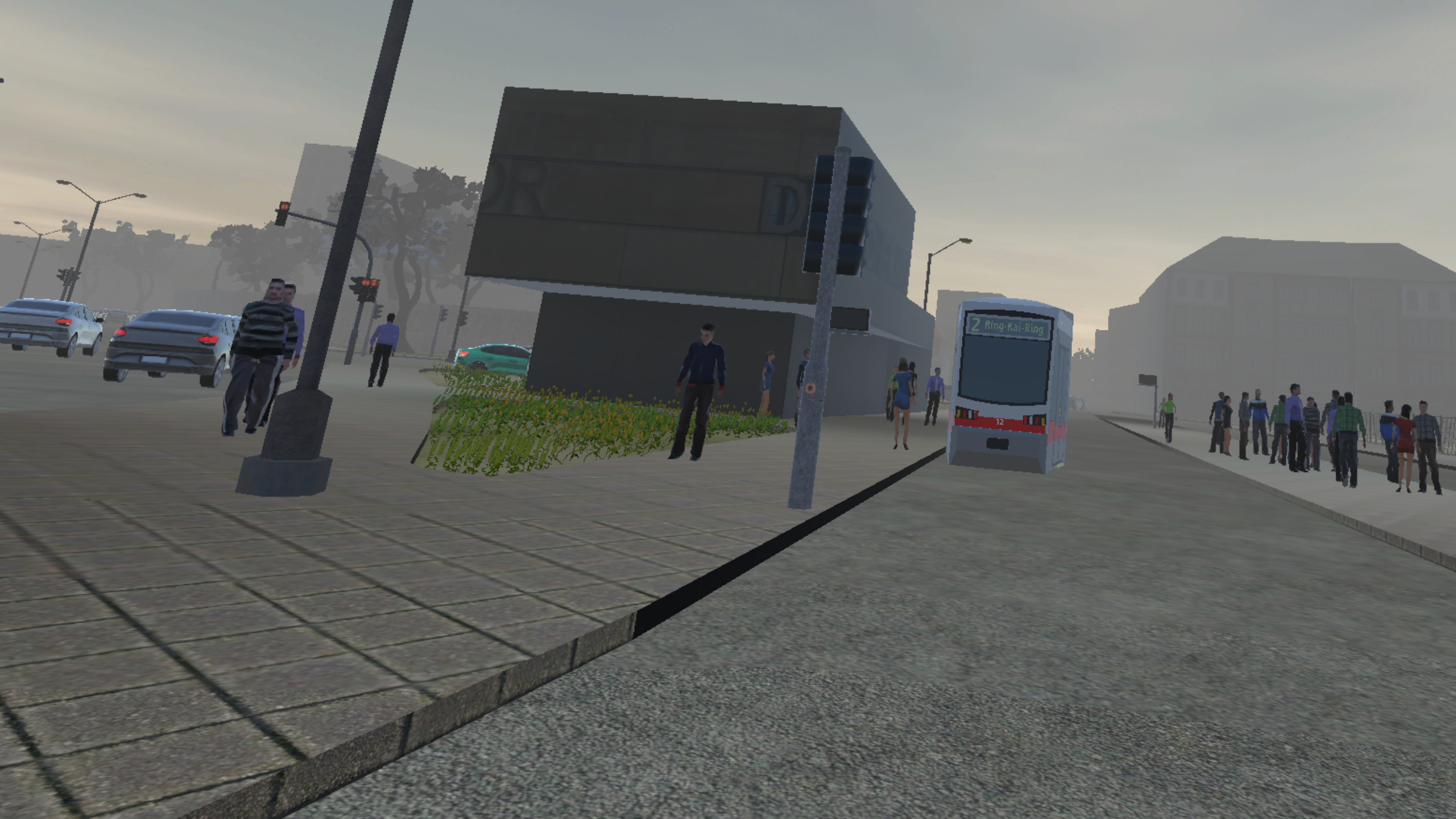}}
\qquad
\subfloat[\Description{Screenshot of the VR simulation. In the view is the underground tram station. Right in front is the lift with a sign indicating that the elevator is broken. To the left, red arrows are hovering along the tram station and pointing towards the alternative working lift.}]{\includegraphics[width=0.3\textwidth]{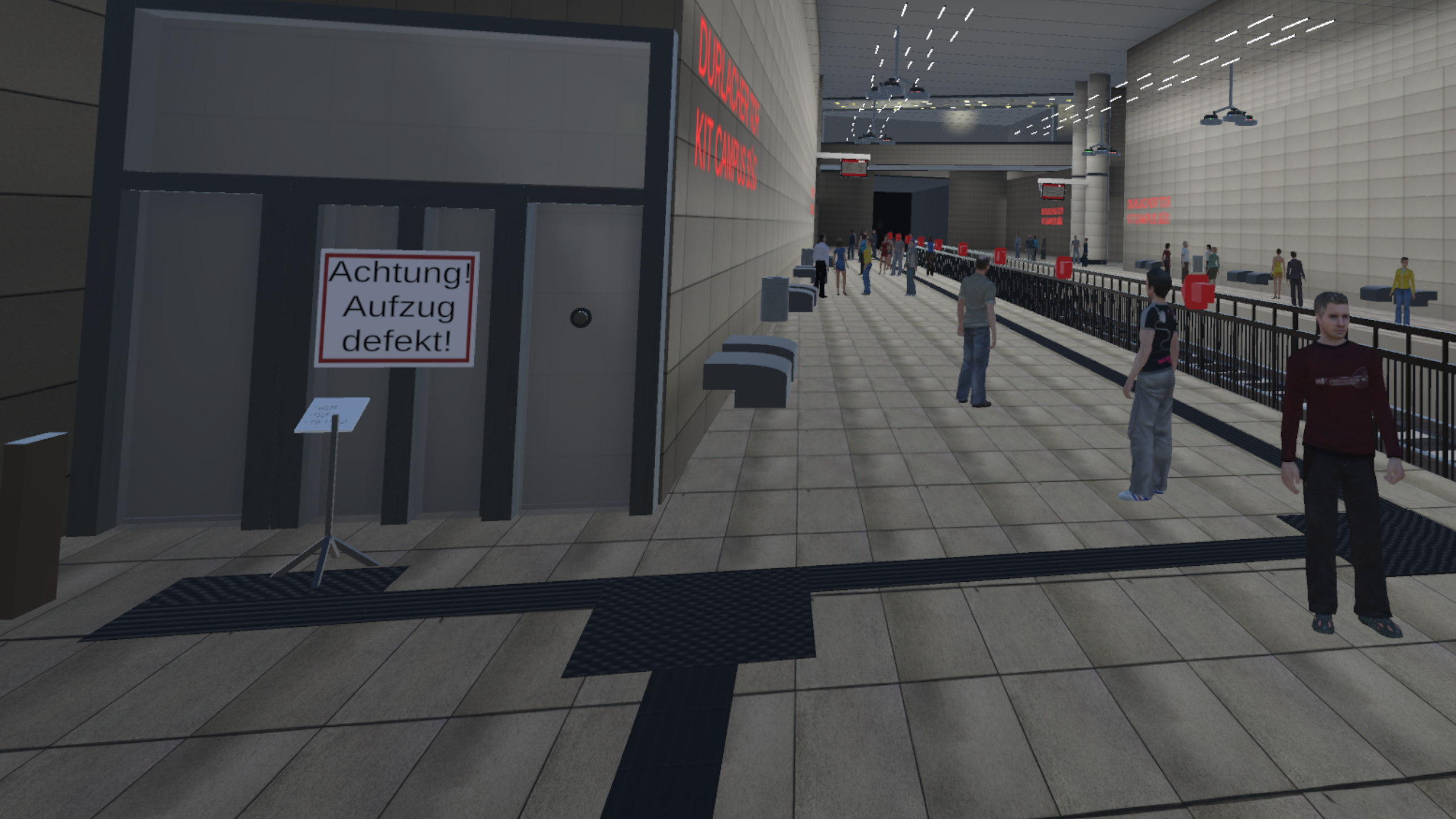}}
  \caption{Screenshots of the VR environment depicting the ground level (left, center) and the underground tram station.}
  \Description{Three horizontally arranged screenshots of the VR environment. The pictures on the left and center show the outside view of the ground level and the underground tram station.}
  \label{fig:screenshots}
\end{teaserfigure}


\maketitle

\section{Introduction and Background}
Disabled people continue to experience barriers in daily life, for example, with respect to mobility and transportation \cite{bezyak2017}, and general access to the built environment \cite{imrie1998}. Viewed through the lens of the social model of disability \cite{owens2015}, there is an understanding that societal structures have a disabling effect on individuals, and the need to address these has been widely recognized in legal frameworks such as the UN Convention on the Rights of Persons with Disabilities \cite{UN_convention_2007}. However, the encouragement of shared responsibility for accessible environments that effectively involves non-disabled people in the identification and removal of barriers remains challenging.


In the past, there have been attempts to raise awareness for disability and foster empathy toward disabled people through disability simulation~\cite{chowdhury2022,nario-redmond2017}. With this method, a non-disabled person is temporarily put in a situation in which they are meant to experience~\textit{"what is it like to have a disability"}{~\cite{flower2007}}. Practically, such experiences have been facilitated in-person, or in the context of immersive media, including Virtual Reality (VR)~\cite{chowdhury2021}. However, the approach of simulating disability has been criticized extensively by advocacy groups, e.g., see ~\cite{todd2017,olson2014}. Additionally, a significant body of academic literature has shown that disability simulation can be ineffective, if not harmful often due to a lack of competence that non-disabled people bring into the simulation (e.g., no experience with the use of assistive devices), which can lead to false assumptions and yield negative feelings such as anger and frustration~\cite{flower2007, nario-redmond2017}. Likewise, there is concern about simulations that equate brief engagement with lived experience of disability~\cite{meinen2023}. 

At the same time, there is evidence that VR can be an effective tool to stimulate reflection, for example, in the context of breastfeeding~\cite{tang2023} or education~\cite{richter2022}. In this work, we therefore want to explore an alternative avenue to disability simulation, adopting an alternative approach in which we highlight barriers rather than implying that we can or should simulate the experience of disability.

We do so through a case study of a VR simulation of a popular traffic hub in Karlsruhe, Germany that is known for its complexity and access barriers. Within the simulation, we leverage visual embellishments~\cite{hicks2019} to focus user attention on the environment rather than on the experience of disability, and we highlight barriers and their removal to promote reflection on what kinds of issues are present in everyday environments, and how they could be addressed through collective action. 
Through this demo, we aim to expand the understanding of how \ac{VR} technology can promote shared responsibility for accessible environments, and we want to provide the foundation for further exploration of technology that is capable of promoting reflection.

\section{A VR Simulation of Barriers in the Built Environment}

The design of the VR simulation aimed to replicate a spot that is familiar to the local community and showcases different barriers that may occur in public spaces. Therefore, we chose the traffic hub \textit{Durlacher Tor}\footnote{\url{https://www.google.com/maps/@49.0089334,8.4170876,20.03z?entry=ttu}} in Karlsruhe, see Figure~\ref{fig:screenshots}. The general flow of the simulation is as follows: After entering VR, the participants find themselves as pedestrians standing on the sidewalk at the junction. They can move freely and interact with the environment. A visual marker highlights the path to the first barrier. After a barrier is reached, the visual guide switches to the next highlighted barrier. This process continues until all three points of interest are visited.

\subsection{System Design}
Here, we give an overview of our system design with focus on  the simulation site and the integration and highlighting of barriers. 

\subsubsection{Identification of Simulation Site and Barriers}
We chose the simulation site because of its complexity: At Durlacher Tor, multiple tram and subway lines meet; there are several bus stops, motorized traffic flows from multiple directions, and pedestrians and cyclists frequently pass by. Additionally, given its close proximity to local businesses and the university campus, crowding occurs multiple times per day. As a result, there are many known access barriers discussed within the local community, making it a worthwhile site to focus on. Additionally, the geographical proximity of our research group to the hub enabled fast development, and allowed us to check the real site and compare it against our simulation frequently.
The total region of the simulation space is approximately $150\times150$ \si{\metre}. 

\subsubsection{Visual Embellishment of (Removal of) Barriers}
\begin{figure*}
\subfloat[Barrier 1: Interrupted tactile guiding strip.\Description{Screenshot of Barrier 1 -- Interrupted tactile guiding strip. The Image depicts a sidewalk with a tactile guiding strip parallel to the road, which runs on the left side. Particle effects above the guiding strip should draw the user's attention toward the barrier.} \label{fig:barrier_1}]{\includegraphics[width=0.3\textwidth]{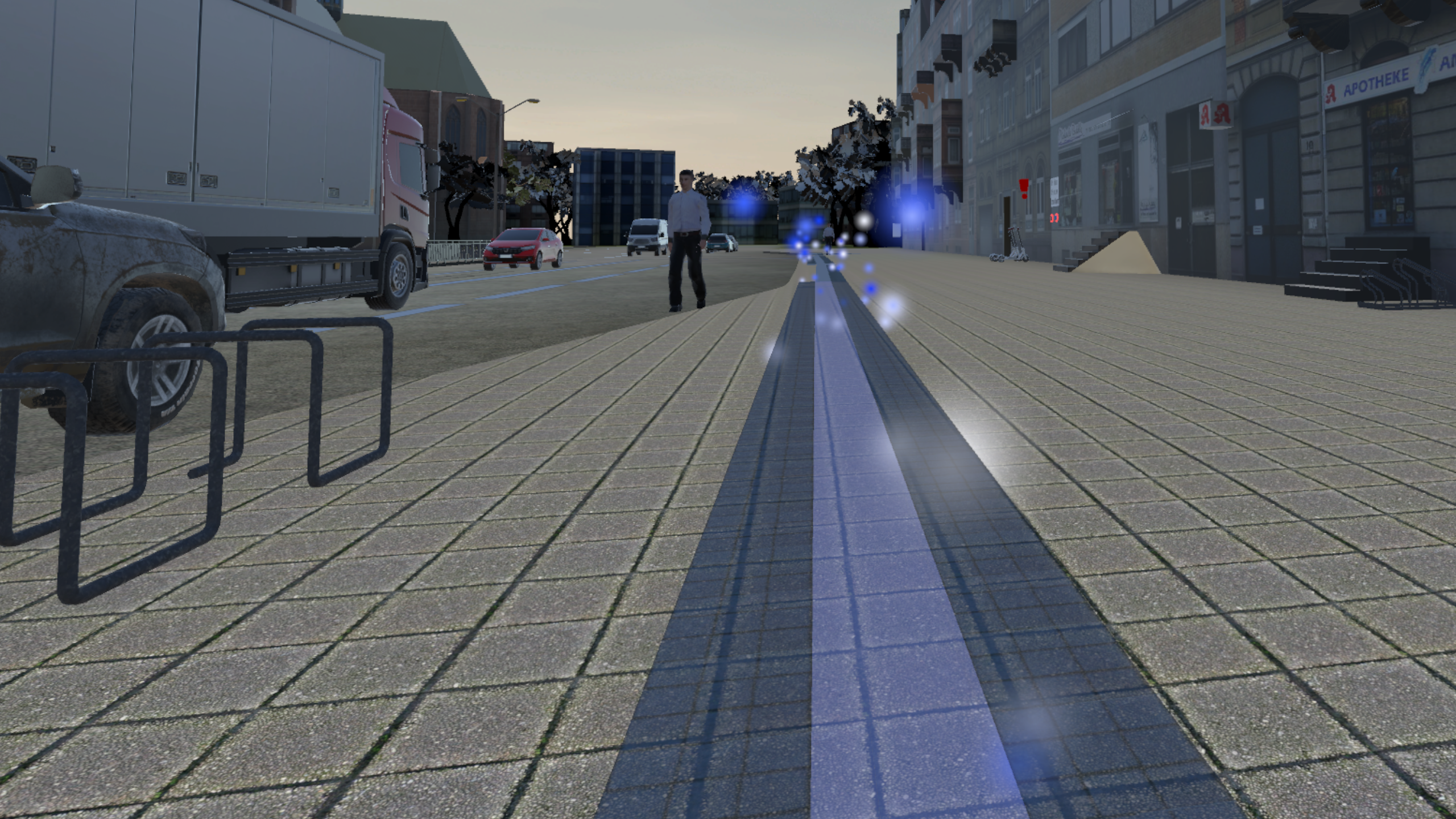}}
\qquad
\subfloat[Barrier 2: Scooters cluttering the sidewalk. \Description{Screenshot of Barrier 2 -- Scooters clustering the sidewalk. The graphic shows a tactile guiding strip on a sidewalk running parallel to the road that is blocked by parking scooters on it. On the ground, blue arrow markings highlight the path towards the barrier. On the right, next to the frontage, the info board with the  exclamation mark hovering is depicted.} \label{fig:barrier_2}]{\includegraphics[width=0.3\textwidth]{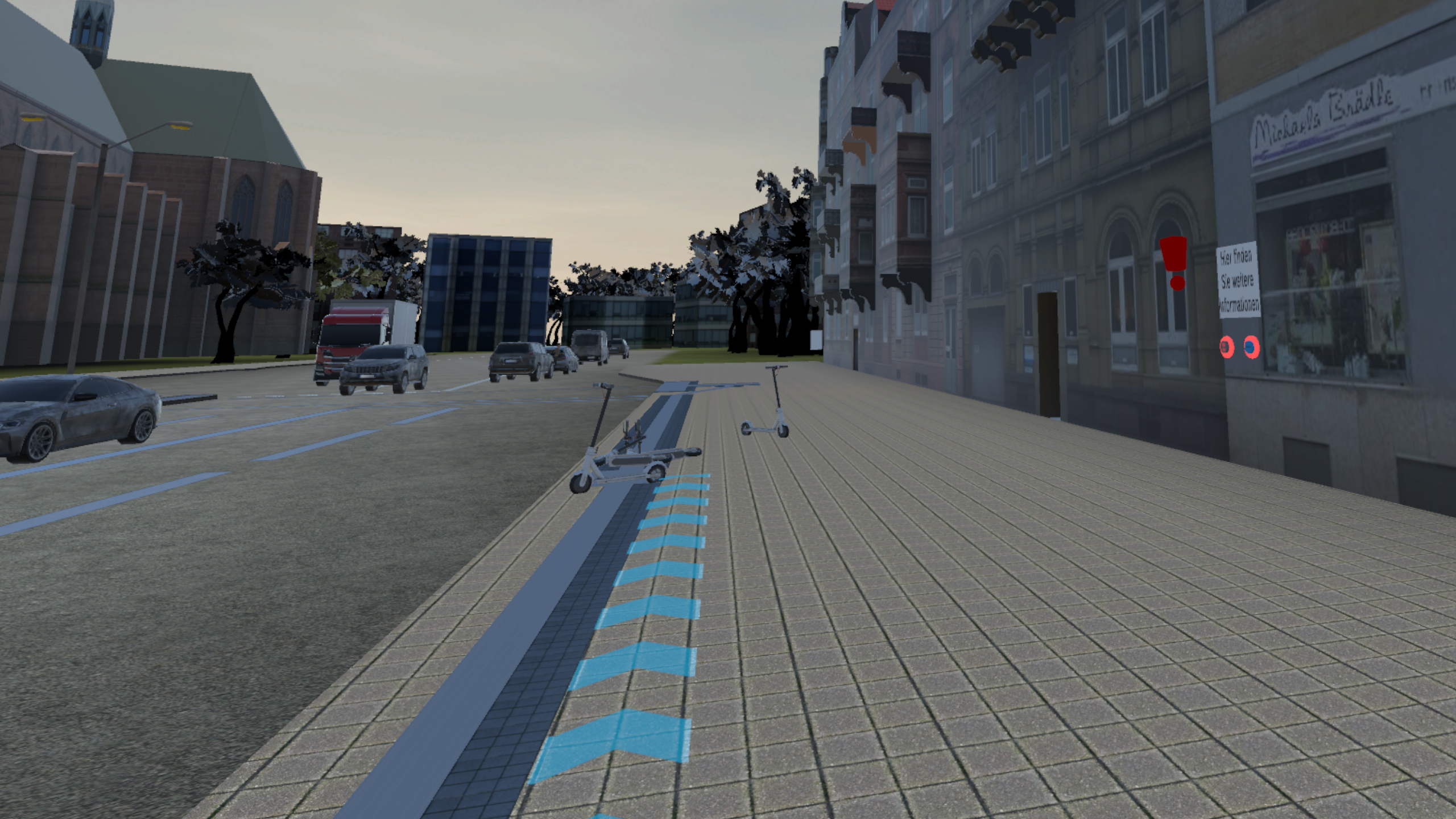}}
\qquad
\subfloat[Barrier 3: Dysfunctional lift. \Description{Screenshot of Barrier 3 - Dysfunctional lift. The picture shows the tram station at the underground level. To the right is the lift with an info sign indicating that the elevator is broken. To the left is an info board providing information about the barrier and an exclamation mark hovering in front of it.} \label{fig:barrier_3}]{\includegraphics[width=0.3\textwidth]{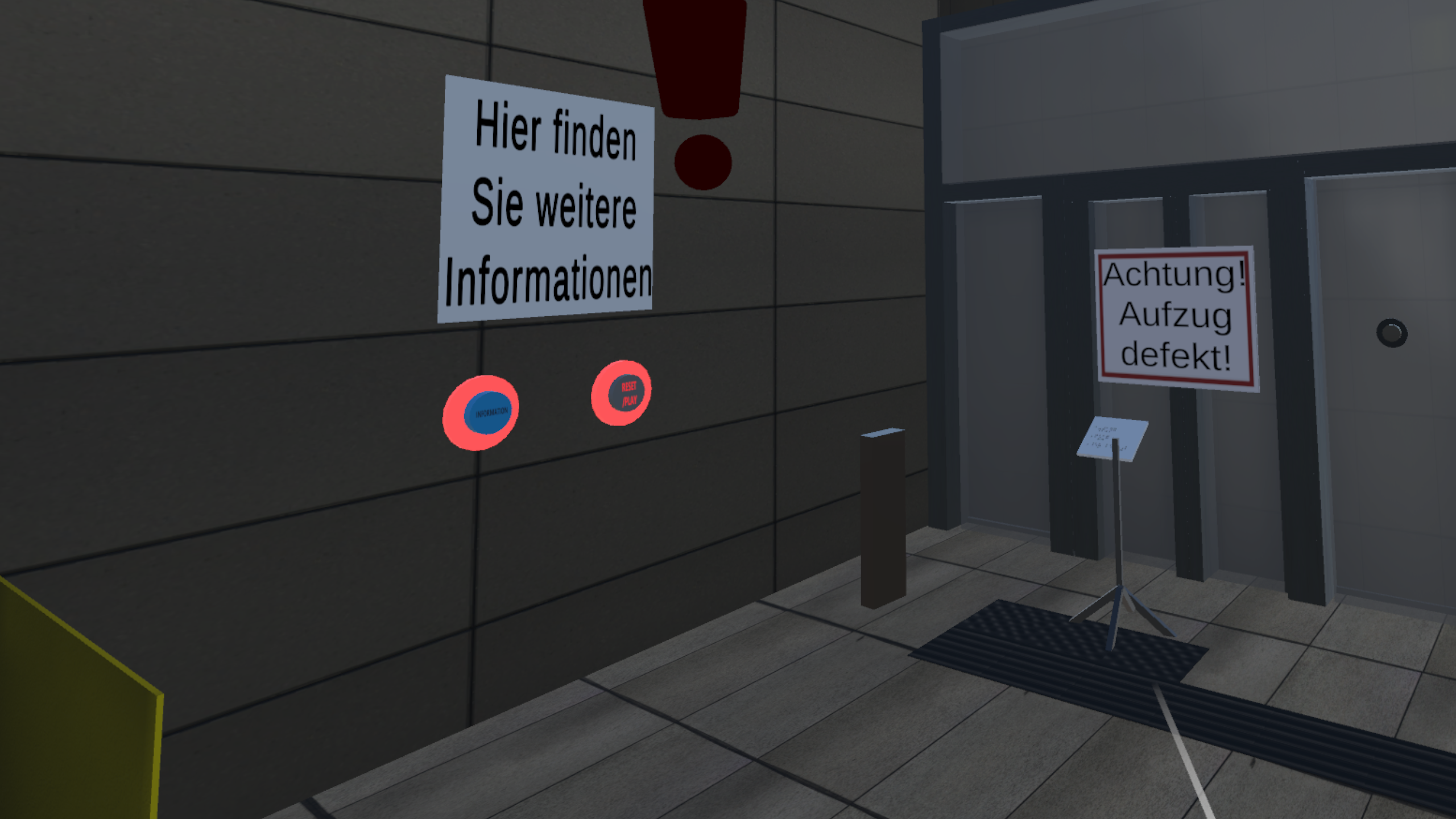}}
  \caption{Screenshots of the barriers and highlighting strategies leveraged in the VR environment.}
  \Description{Horizontally arranged screenshots of the VR simulation depicting three barriers. Each screenshot visualizes a strategy to highlight barriers in the environment.}
  \label{fig:barriers}
\end{figure*}

To spark the users' reflection processes, the simulation hints the user toward three different barriers in the virtual environment (see Figure~\ref{fig:barriers}), and we employ visual embellishments to alert users to their existence: Each barrier is marked with an exclamation mark hovering above it. When the user is in proximity to a barrier, a particle effect starts playing to draw the user's attention toward it. 
The path towards the barriers is visualized on the ground to guide the player to the barriers.

\paragraph{Barrier 1: Abruptly ending tactile guiding strip} On the sidewalk at ground level, there is a tactile guiding strip that unexpectedly ends without providing any information for a relying person who is blind or has low vision where to go (Figure~\ref{fig:barrier_1}). Once the barrier was encountered, the simulation places an additional piece of tactile paving to improve access.

\paragraph{Barrier 2: Cluttered sidewalk} The sidewalk is cluttered by scooters (see Figure~\ref{fig:barrier_2}), which is a challenge for a range of people, including wheelchair users and people who are blind or have low vision. When the user is in proximity, an animation starts, which moves the scooters to the side so that the guide strip and the sidewalk can be safely navigated.

\paragraph{Barrier 3: Broken elevator} One of the elevators to reach the underground station is broken (see~\ref{fig:barrier_3}), posing a barrier for people with limited mobility. When the user reaches the broken elevator, a sign informs them that it does not work and that they are required to take a different one on the other side of the station. Additionally, moving arrows indicate the way to the alternative elevator.

We selected the initial barriers based on the literature on pedestrian navigation safety~\cite{fuss2022, neumann2022} and internal discussions.
In future work, we want to explore integration of additional barriers that can result from dynamic situations (e.g., crowding), and also include different types of barriers apart from mobility-related ones that can be hindrances on other levels, e.g., sensory barriers, such as crowding and noise, or motor barriers, e.g., intricate input modalities for ticket machines.
With this approach, we aim to broaden the perspective on how people experience the world and what constitutes an access barrier.

\subsection{Technical Implementation}
The VR simulation is built using the game engine Unity 3D 2022\footnote{\url{https://unity.com}}. To design the virtual environment, we used 3D geographic data provided by the city administration Geoportal Karlsruhe\footnote{\url{http://geoportal.karlsruhe.de/}}. For the simulation of pedestrians and the traffic, we employed the Mobile Traffic System\footnote{\url{https://gleygames.com/traffic-system/}} package, which handles the pathfinding of the road users, the traffic light logic and car overtaking prioritization. To provide a versatile experience, the simulation includes $107$ 3D models of pedestrians and  $19$ different car types. 
We used the original schedule for public transport, resulting in public transport arriving approximately every $5{-}10$~\si{\minute}.
For the VR interface, we used the OpenXR framework in conjunction with the Interaction Toolkit and employed the standard controller inputs: left joystick for continuous movement, right joystick for discrete rotation in steps of \SI{45}{\degree} and the trigger buttons to interact with the environment.
The application runs on a desktop PC with NVIDIA GeForce RTX 3070 graphics, Intel i5 11500 processor, and 16 GB RAM. 
As the VR device, we use the Meta Quest 3 running as a client device via Oculus Link.

\section{Outlook and On-Site Demonstration at MuC 2024}
In this section, we give an overview of the anticipated setup at the conference and details of the demonstration process. Additionally, we reflect upon the experience that we hope participants will have when interacting with the demo, and we provide guiding questions that we want to explore together with participants.

\subsection{Demo Setup and Requirements}
We will present the VR experience at the venue. Participants will be invited to try out the demo or observe while the authors demonstrate the application. We anticipate that each participant will need to spend about five minutes in VR. For the installation, we will provide a desktop PC and a VR headset. On-site, we will require a $2\times3~\si{\metre}$ space with a large (35 - 50 inch) display and two tables with three chairs. 

\subsection{Participant Experience and Potential for Joint Reflection at the Conference}
With this demo, we aim to spark discussions around inclusive design and how VR can leverage a reflection on barriers in public spaces while avoiding harmful effects yielded by disability simulations. We especially invite visitors with disabilities, as well as HCI researchers involved in interaction design for accessibility, to explore the demo and join us in reflecting upon the following issues: What access barriers exist in the built environment, and which ones are commonly overlooked by non-disabled people? What insights can be gained from a VR simulation with respect to barriers? In which way does knowledge transfer to other settings within the simulation and in the real world? What are respect- but insightful ways of communicating barriers in VR, and in how far does such a simulation address shortcomings of disability simulation? And finally, are there risks in our approach to simulating barriers? 




\bibliographystyle{ACM-Reference-Format}
\bibliography{main.bib}


\begin{thebibliography}{16}


\ifx \showCODEN    \undefined \def \showCODEN     #1{\unskip}     \fi
\ifx \showDOI      \undefined \def \showDOI       #1{#1}\fi
\ifx \showISBNx    \undefined \def \showISBNx     #1{\unskip}     \fi
\ifx \showISBNxiii \undefined \def \showISBNxiii  #1{\unskip}     \fi
\ifx \showISSN     \undefined \def \showISSN      #1{\unskip}     \fi
\ifx \showLCCN     \undefined \def \showLCCN      #1{\unskip}     \fi
\ifx \shownote     \undefined \def \shownote      #1{#1}          \fi
\ifx \showarticletitle \undefined \def \showarticletitle #1{#1}   \fi
\ifx \showURL      \undefined \def \showURL       {\relax}        \fi
\providecommand\bibfield[2]{#2}
\providecommand\bibinfo[2]{#2}
\providecommand\natexlab[1]{#1}
\providecommand\showeprint[2][]{arXiv:#2}

\bibitem[Bezyak et~al\mbox{.}(2017)]%
        {bezyak2017}
\bibfield{author}{\bibinfo{person}{Jill~L. Bezyak}, \bibinfo{person}{Scott~A.
  Sabella}, {and} \bibinfo{person}{Robert~H. Gattis}.}
  \bibinfo{year}{2017}\natexlab{}.
\newblock \showarticletitle{Public Transportation: An Investigation of Barriers
  for People With Disabilities}.
\newblock \bibinfo{journal}{\emph{Journal of Disability Policy Studies}}
  \bibinfo{volume}{28}, \bibinfo{number}{1} (\bibinfo{date}{June}
  \bibinfo{year}{2017}), \bibinfo{pages}{52--60}.
\newblock
\showISSN{1044-2073, 1538-4802}
\urldef\tempurl%
\url{https://doi.org/10.1177/1044207317702070}
\showDOI{\tempurl}


\bibitem[Chowdhury et~al\mbox{.}(2021)]%
        {chowdhury2021}
\bibfield{author}{\bibinfo{person}{Tanvir~Irfan Chowdhury},
  \bibinfo{person}{Sharif Mohammad~Shahnewaz Ferdous}, {and}
  \bibinfo{person}{John Quarles}.} \bibinfo{year}{2021}\natexlab{}.
\newblock \showarticletitle{VR Disability Simulation Reduces Implicit Bias
  Towards Persons With Disabilities}.
\newblock \bibinfo{journal}{\emph{IEEE Transactions on Visualization and
  Computer Graphics}} \bibinfo{volume}{27}, \bibinfo{number}{6}
  (\bibinfo{date}{June} \bibinfo{year}{2021}), \bibinfo{pages}{3079--3090}.
\newblock
\showISSN{1077-2626, 1941-0506, 2160-9306}
\urldef\tempurl%
\url{https://doi.org/10.1109/TVCG.2019.2958332}
\showDOI{\tempurl}


\bibitem[Chowdhury and Quarles(2022)]%
        {chowdhury2022}
\bibfield{author}{\bibinfo{person}{Tanvir~Irfan Chowdhury} {and}
  \bibinfo{person}{John Quarles}.} \bibinfo{year}{2022}\natexlab{}.
\newblock \showarticletitle{A Wheelchair Locomotion Interface in a VR
  Disability Simulation Reduces Implicit Bias}.
\newblock \bibinfo{journal}{\emph{IEEE Transactions on Visualization and
  Computer Graphics}} \bibinfo{volume}{28}, \bibinfo{number}{12}
  (\bibinfo{date}{Dec.} \bibinfo{year}{2022}), \bibinfo{pages}{4658--4670}.
\newblock
\showISSN{1077-2626, 1941-0506, 2160-9306}
\urldef\tempurl%
\url{https://doi.org/10.1109/TVCG.2021.3099115}
\showDOI{\tempurl}


\bibitem[Flower et~al\mbox{.}(2007)]%
        {flower2007}
\bibfield{author}{\bibinfo{person}{Ashley Flower}, \bibinfo{person}{Matthew~K.
  Burns}, {and} \bibinfo{person}{Nicole~A. {Bottsford-Miller}}.}
  \bibinfo{year}{2007}\natexlab{}.
\newblock \showarticletitle{Meta-Analysis of Disability Simulation Research}.
\newblock \bibinfo{journal}{\emph{Remedial and Special Education}}
  \bibinfo{volume}{28}, \bibinfo{number}{2} (\bibinfo{date}{March}
  \bibinfo{year}{2007}), \bibinfo{pages}{72--79}.
\newblock
\showISSN{0741-9325, 1538-4756}
\urldef\tempurl%
\url{https://doi.org/10.1177/07419325070280020601}
\showDOI{\tempurl}


\bibitem[FUSS(2022)]%
        {fuss2022}
\bibfield{author}{\bibinfo{person}{e.V. FUSS}.}
  \bibinfo{year}{2022}\natexlab{}.
\newblock \bibinfo{booktitle}{\emph{Gest{\"o}rte Mobilit{\"a}t: Daten Und
  Fakten Zu E-Scootern \& Co. Auf Berliner Gehwegen}}.
\newblock \bibinfo{type}{{T}echnical {R}eport}.
\newblock


\bibitem[Hicks et~al\mbox{.}(2019)]%
        {hicks2019}
\bibfield{author}{\bibinfo{person}{Kieran Hicks}, \bibinfo{person}{Kathrin
  Gerling}, \bibinfo{person}{Patrick Dickinson}, {and} \bibinfo{person}{Vero
  Vanden~Abeele}.} \bibinfo{year}{2019}\natexlab{}.
\newblock \showarticletitle{Juicy Game Design: Understanding the Impact of
  Visual Embellishments on Player Experience}. In
  \bibinfo{booktitle}{\emph{Proceedings of the Annual Symposium on
  Computer-Human Interaction in Play}}. \bibinfo{publisher}{ACM},
  \bibinfo{address}{Barcelona Spain}, \bibinfo{pages}{185--197}.
\newblock
\showISBNx{978-1-4503-6688-5}
\urldef\tempurl%
\url{https://doi.org/10.1145/3311350.3347171}
\showDOI{\tempurl}


\bibitem[Imrie and Kumar(1998)]%
        {imrie1998}
\bibfield{author}{\bibinfo{person}{Rob Imrie} {and} \bibinfo{person}{Marion
  Kumar}.} \bibinfo{year}{1998}\natexlab{}.
\newblock \showarticletitle{Focusing on Disability and Access in the Built
  Environment}.
\newblock \bibinfo{journal}{\emph{Disability \& Society}} \bibinfo{volume}{13},
  \bibinfo{number}{3} (\bibinfo{date}{June} \bibinfo{year}{1998}),
  \bibinfo{pages}{357--374}.
\newblock
\showISSN{0968-7599, 1360-0508}
\urldef\tempurl%
\url{https://doi.org/10.1080/09687599826687}
\showDOI{\tempurl}


\bibitem[Meinen(2023)]%
        {meinen2023}
\bibfield{author}{\bibinfo{person}{Lisanne~E. Meinen}.}
  \bibinfo{year}{2023}\natexlab{}.
\newblock \showarticletitle{Share the Experience, Don't Take It: Toward
  Attunement with Neurodiversity in Videogames}.
\newblock \bibinfo{journal}{\emph{Games and Culture}} \bibinfo{volume}{0},
  \bibinfo{number}{0} (\bibinfo{year}{2023}),
  \bibinfo{pages}{15554120221149538}.
\newblock
\urldef\tempurl%
\url{https://doi.org/10.1177/15554120221149538}
\showDOI{\tempurl}
\showeprint{https://doi.org/10.1177/15554120221149538}


\bibitem[{Nario-Redmond} et~al\mbox{.}(2017)]%
        {nario-redmond2017}
\bibfield{author}{\bibinfo{person}{M.R. {Nario-Redmond}}, \bibinfo{person}{D.
  Gospodinov}, {and} \bibinfo{person}{A A.Cobb}.}
  \bibinfo{year}{2017}\natexlab{}.
\newblock \showarticletitle{Crip for a Day: The Unintended Negative
  Consequences of Disability Simulations}.
\newblock \bibinfo{journal}{\emph{Rehabilitation Psychology}}
  \bibinfo{volume}{62}, \bibinfo{number}{3} (\bibinfo{year}{2017}),
  \bibinfo{pages}{324--333}.
\newblock
\urldef\tempurl%
\url{https://doi.org/10.1037/rep0000127}
\showDOI{\tempurl}


\bibitem[Nations(2006)]%
        {UN_convention_2007}
\bibfield{author}{\bibinfo{person}{United Nations}.}
  \bibinfo{year}{2006}\natexlab{}.
\newblock \showarticletitle{Convention on the Rights of Persons with
  Disabilities}.
\newblock \bibinfo{journal}{\emph{Treaty Series}}  \bibinfo{volume}{2515}
  (\bibinfo{date}{Dec.} \bibinfo{year}{2006}), \bibinfo{pages}{3}.
\newblock


\bibitem[Neumann(2022)]%
        {neumann2022}
\bibfield{author}{\bibinfo{person}{Peter Neumann}.}
  \bibinfo{year}{2022}\natexlab{}.
\newblock \bibinfo{title}{Blindenverein reicht Klage ein: E-Scooter sollen von
  den Gehwegen verschwinden}.
\newblock
  \bibinfo{howpublished}{https://www.berliner-zeitung.de/mensch-metropole/blindenverein-reicht-klage-ein-e-scooter-sollen-von-gehwegen-verschwinden-berlin-lime-voi-absv-li.273229}.
\newblock


\bibitem[Olson(2014)]%
        {olson2014}
\bibfield{author}{\bibinfo{person}{Olson}.} \bibinfo{year}{2014}\natexlab{}.
\newblock \bibinfo{title}{How Disability Simulations Promote Damaging
  Stereotypes}.
\newblock
  \bibinfo{howpublished}{https://nfb.org/sites/default/files/images/nfb/publications/bm/bm14/bm1401/bm140107.htm}.
\newblock


\bibitem[Owens(2015)]%
        {owens2015}
\bibfield{author}{\bibinfo{person}{Janine Owens}.}
  \bibinfo{year}{2015}\natexlab{}.
\newblock \showarticletitle{Exploring the Critiques of the Social Model of
  Disability: The Transformative Possibility of Arendt's Notion of Power}.
\newblock \bibinfo{journal}{\emph{Sociology of Health \& Illness}}
  \bibinfo{volume}{37}, \bibinfo{number}{3} (\bibinfo{date}{March}
  \bibinfo{year}{2015}), \bibinfo{pages}{385--403}.
\newblock
\showISSN{0141-9889, 1467-9566}
\urldef\tempurl%
\url{https://doi.org/10.1111/1467-9566.12199}
\showDOI{\tempurl}


\bibitem[Richter et~al\mbox{.}(2022)]%
        {richter2022}
\bibfield{author}{\bibinfo{person}{Eric Richter}, \bibinfo{person}{Isabell
  Hu{\ss}ner}, \bibinfo{person}{Yizhen Huang}, \bibinfo{person}{Dirk Richter},
  {and} \bibinfo{person}{Rebecca Lazarides}.} \bibinfo{year}{2022}\natexlab{}.
\newblock \showarticletitle{Video-Based Reflection in Teacher Education:
  Comparing Virtual Reality and Real Classroom Videos}.
\newblock \bibinfo{journal}{\emph{Computers \& Education}}
  \bibinfo{volume}{190} (\bibinfo{date}{Dec.} \bibinfo{year}{2022}),
  \bibinfo{pages}{104601}.
\newblock
\showISSN{03601315}
\urldef\tempurl%
\url{https://doi.org/10.1016/j.compedu.2022.104601}
\showDOI{\tempurl}


\bibitem[Tang et~al\mbox{.}(2023)]%
        {tang2023}
\bibfield{author}{\bibinfo{person}{Kymeng Tang}, \bibinfo{person}{Kathrin
  Gerling}, \bibinfo{person}{Vero Vanden~Abeele}, \bibinfo{person}{Luc Geurts},
  {and} \bibinfo{person}{Maria Aufheimer}.} \bibinfo{year}{2023}\natexlab{}.
\newblock \showarticletitle{Playful Reflection: Impact of Gamification on a
  Virtual Reality Simulation of Breastfeeding}. In
  \bibinfo{booktitle}{\emph{Proceedings of the 2023 CHI Conference on Human
  Factors in Computing Systems}}. \bibinfo{publisher}{ACM},
  \bibinfo{address}{Hamburg Germany}, \bibinfo{pages}{1--13}.
\newblock
\showISBNx{978-1-4503-9421-5}
\urldef\tempurl%
\url{https://doi.org/10.1145/3544548.3580751}
\showDOI{\tempurl}


\bibitem[Todd(2017)]%
        {todd2017}
\bibfield{author}{\bibinfo{person}{Zara Todd}.}
  \bibinfo{year}{2017}\natexlab{}.
\newblock \bibinfo{title}{Human Rights Education and Disability Simulation
  Exercises -- Not a Match Made in Heaven - Coyote Magazine - Pjp-Eu.Coe.Int}.
\newblock
  \bibinfo{howpublished}{https://pjp-eu.coe.int/en/web/coyote-magazine/hre-and-disability-simulation}.
\newblock


\end{thebibliography}






\end{document}